\documentclass[%
reprint,
superscriptaddress,
nobibnotes,
amsmath,amssymb,
aps,
pra,
longbibliography
floatfix
]{revtex4-2}
\usepackage{graphicx}  
\usepackage{dcolumn}   
\usepackage{bm}        
\usepackage{amssymb}   
\usepackage[colorlinks=true]{hyperref}
\usepackage{subfigure}
\usepackage{xcolor}
\usepackage{footmisc}
\usepackage[normalem]{ulem}
\usepackage{filecontents}
\DeclareMathOperator{\sgn}{sgn}

\newcommand{\mw}{\textcolor{black}}
\newcommand{\zs}{\textcolor{black}}
\begin{document}
\preprint{APS/123-QED}

\title{Optimality and annealing path planning of dynamical analog solvers }
\author{Shu Zhou}
\author{K. Y. Michael Wong}
\email{phkywong@ust.hk}
\author{Juntao Wang}
\affiliation{Department of Physics, The Hong Kong University of Science and Technology, Hong Kong SAR, China.}
\author{David Shui Wing Hui}
\author{Daniel Ebler}
\author{Jie Sun}
\affiliation{Theory Lab, Central Research Institute, 2012 Labs, Huawei Technologies Co. Ltd., Hong Kong SAR, China.}
\begin{abstract}

    Recently proposed analog solvers based on dynamical systems, such as Ising machines, are promising platforms for large-scale combinatorial optimization. Yet, given the heuristic nature of the field, there is very limited insight on optimality guarantees of the solvers, as well as how parameter schedules shape dynamics and outcomes. Here, we develop a dynamical mean-field framework to analyze Ising-machine dynamics for finding the ground state energy of the Sherrington-Kirkpatrick (SK) model of spin glasses and identify mechanisms that enable rapid convergence to \zs{provably}  near-optimal energies. For a fixed target energy density $E_c$, we show that solutions are typically reached within $O(1)$ matrix vector multiplications, indicating constant time complexity. We further delineate theoretical limitations arising from different parameter-scheduling trajectories and demonstrate a pronounced benefit of temperature-only annealing for the Coherent Ising Machine. Building on these insights, we propose a general framework for designing optimized parameter schedules, thereby improving the practical effectiveness of Ising machines for complex optimization tasks. The superior performance of the dynamical solvers is illustrated by the attainment of the ground state energy of the SK model.
\end{abstract}

\maketitle

\section{Introduction}

Discrete optimization problems that can be formulated as Ising Hamiltonians have found broad applicability across various real-world domains, including information processing \cite{LeVinee17052895}, route planning \cite{Bao9564593}, protein folding\cite{Ooka2023},  and machine learning \cite{Bryngelson1987-yp}.  The problem objective can be written as
\begin{equation}
    \min_{\sigma\in\{-1,+1\}^N}\left(-\frac{1}{2}\sum_{ij} \sigma_iJ_{ij}\sigma_j\right),\label{eq:ising_problem}
\end{equation}
where the optimization variables are Ising spins $\sigma\in\{\pm1\}^N$, which  are coupled through the problem matrix $[J_{ij}]_{N\times N}$.
The search for the spin configurations that minimize this Hamiltonian is widely regarded as computationally intractable for classical computers \cite{FBarahona_1982,arora2005non}.

Recently, with the development of new computational paradigms such as quantum and quantum-inspired computing, there is an upsurge of interest in non-Von-Neumann computing schemes to solve discrete optimization problems efficiently. Concretely, such schemes simulate the dynamics of classical or quantum variables, referred to as (analog) spins, and map them to their Ising counterparts via nonlinear activation functions. The Ising energy after mapping is referred to as the decoded energy. Various types of such devices have emerged, leveraging different physical spin hardware \cite{Mohseni2022}. Examples include quantum annealers based on superconducting circuits \cite{PhysRevE.58.5355, Johnson2011QuantumAW, Boixo_2014, Hauke_2020}, digital systems utilizing special-purpose classical processors \cite{PhysRevX.5.021027, Chou2019, 7350099, Aramon_2019, Yamamoto2020, Leleu2021-cz}, and optical Ising machines employing pulse lasers \cite{Marandi_2014, Yamamoto2017, PhysRevA.88.063853}. Among these, the Coherent Ising Machine (CIM) stands out as a notable example, demonstrating superior scalability and efficiency \cite{inagaki2016coherent, doi:10.1126/sciadv.abh0952}.  Numerous CIM variations~\cite{Order-of-magnitude, Mohseni2022, Bohm2019-pi, Tiunov:19, Goto2021} and theoretical studies~\cite{Leleu_2019, Yamamura2023, Zhou_2025} have recently been proposed. Yet, despite advances in understanding steady states, the design of dynamical algorithms remains largely intuitive. The annealing schedule, crucial for guiding the system from high to low entropy via gain parameter adjustment~\cite{PhysRevA.88.063853}, is often obstructed by metastable states~\cite{Yamamura2023} and gaps in spin distributions~\cite{Zhou_2025} that restrict necessary spin flips. To circumvent these convergence issues, a technique known as Chaotic Amplitude Control (CAC)~\cite{Leleu_2019} was introduced, utilizing sophisticated parameter tuning and specially designed dynamics, and has shown promise in numerical scaling experiments~\cite{Leleu2021-cz}. Nevertheless, the systematic scaling of Ising machine precision with network size remains unknown from both theoretical and experimental perspectives. This highlights the necessity for a systematic dynamical framework to evaluate performance—specifically, to determine how solution quality correlates with the machine's dynamical evolution.

\zs{In this paper, we derive annealing schedules that are optimal within the dynamics of a given machine. Although previous studies have demonstrated that noise  can be beneficial in various settings \cite{bohm2022noise,kanao2022simulated,lee2025noise,goto2026edge}, the most widely adopted annealing protocol still varies only the linear gain $a$ while maintaining weak noise throughout the evolution~\cite{Mohseni2022,Leleu_2019,PhysRevA.88.063853,Yamamura2023}. Our analysis shows that, for typical platforms such as CIMs, such gain-only schedules are generally suboptimal and, in particular, do not lead to the fixed-accuracy polynomial-time scaling \mw{as revealed in recent algorithmic developments \cite{Alaoui_2020,Alaoui_2021,Montanari_2019}}. This highlights the importance of coordinated parameter scheduling, most notably involving temperature, for achieving robust and scalable optimization. }

\zs{A key insight is the central role played by the density of near-zero soft spins during the dynamics. To quantify this effect, we introduce the concept of an effective gap, which connects the DMFT description to experimentally accessible dynamical activity and provides a practical criterion for schedule design. For CIMs, our analysis identifies \mw{a} superior schedule that anneals only the noise, in contrast to common gain-only practice~\cite{Mohseni2022,Leleu_2019,PhysRevA.88.063853,Yamamura2023}. We further extend the framework to broader Ising-machine architectures and identify cases in which the theory instead recommends the gain-only practice during operation. }

Our study of how the dynamics of Ising machines determines the accuracy of the output in optimization tasks is based on the search for ground state energy of the Sherrington-Kirkpatrick (SK) model~\cite{Sherrington1975, Kirkpatrick1977} of spin glasses. As a prototype of an optimization problem, the complex energy landscape of the SK model is typical of general combinatorial optimization problems~\cite{Boettcher2004, Kim2007, Palassini_2008}, and the numerical value of its ground state energy is well known for benchmarking~\cite{Boettcher2004, Kim2007, Palassini_2008}. Using the dynamical mean-field theory developed by Sompolinsky {\it et al.}~\cite{Sompolinsky_1981, Sompolinsky_1982}, we model the CIM using a soft-spin Ginzburg–Landau mean-field dynamics. This yields time-dependent expressions for the decoded energy across several Ising-machine protocols. Using these expressions, together with Monte Carlo and numerical experiments, we find that the decoded energy reaches the target value in an average of $O(1)$ steps of matrix vector multiplication, each with $O(N^2)$ operations. Furthermore, we found that the required precision of the CIM dynamics {at finite time} scales with the system size in the same manner as the finite-size scaling of the ground state energy.

The significance of this result is highlighted by considering recent algorithmic developments, in which message-passing methods produce solutions that are arbitrarily close to the ground state with time complexity $C(\epsilon)N^2$ where $\epsilon$ is the fractional discrepancy with the ground state  \cite{Alaoui_2020,Alaoui_2021,Montanari_2019}. This demonstrates that despite the NP-hardness of Ising optimization in the worst case, near-optimal solutions are attainable in polynomial time for mean-field models.
These schemes rely on oracle access to the Parisi functional minimizer. Although this minimizer has been extensively studied within the context of the SK model, such a dependency may limit the applicability of these methods to more general practical scenarios
\cite{Auffinger_2014,Talagrand_2011}. \zs{In this context, our results successfully capture the desired $N$-scaling behavior in physical Ising machines, thereby translating this theoretical advantage into experimentally relevant dynamics.} 

The remainder of this paper is organized as follows: In Section 2, we introduce our methodology and derive the dynamic mean-field expression for the decoded Ising energy. Section 3 presents an analysis of the system's time complexity, raising the fundamental question {on} the accessibility of a target energy for a given Ising machine. To address this, Sections 4 and 5 unveil the soft-hard spin dichotomy mechanism and identify the dynamical barrier to evolution, which we define as the `effective gap'. Based on these findings, we propose an optimal annealing path {through temperature annealing} for CIMs that {outperforms} conventional approaches {via gain annealing}. In Section 6, we demonstrate that this framework can yield different optimal annealing paths for different systems and illustrate the phenomenon of barrier exposure. Finally, in Section 8, utilizing the techniques developed in previous sections, we answer the question raised in Section 3 through asymptotic numerical analysis, providing evidence that Ising machines can access the ground state with an error margin
    {$\epsilon\ll 1$} within polynomial time.

\section{Dynamical Mean-field Theory (DMFT)}

 {We first consider the example of CIM, which is governed by the following dynamical equation:}
\begin{equation}
    \frac{dx_i}{dt} = -x_{i}^{3} + a x_{i}+ \xi \sum_{j=1, j \neq i}^{N} J_{ij } x_{j} + h_i+ \zeta_i.
    \label{eq:CIM}
\end{equation}
where $x_i$ represents the state of the $i$-th spin, $a$ is a gain coefficient, and $\xi$ denotes the coupling strength~\cite{PhysRevA.88.063853}. \zs{Unless otherwise stated, we set $\xi =1$ by absorbing it into the definition of the gain parameter.} The noise term $\zeta_i$ is modeled as white noise characterized by a temperature $T$, satisfying $\langle \zeta_i(t) \rangle = 0$ and $\langle \zeta_i(t) \zeta_j(t') \rangle = 2T\delta_{ij}\delta(t-t')$. Additionally, we focus on Ising problems defined by the SK model, where the couplings $J_{ij}$ are drawn from a Gaussian distribution with zero mean and variance $J^2/N$~\cite{Sherrington1975, Kirkpatrick1977}. $h_i$
is an external field set to be 0.

Such a system shares the mathematical structure of the time-dependent soft-spin Ginzburg–Landau model, analyzed by Sompolinsky and Zippelius to characterize the spin-glass phase of the SK model in equilibrium~\cite{Sompolinsky_1981, Sompolinsky_1982}. Referred to as the dynamical mean-field theory (DMFT), this methodology traces the evolution of a generating functional to extract the dynamical variables of the system~\cite{Martin_1973, Chow_2010, Grzetic_2014}. It was later applied to the dynamics of models such as the $p$-spin glass~\cite{Cugliandolo_1993}.

The Ising configuration is read out from the signs of the analog spins $s_i = \sgn(x_i)$. The associated Ising energy is then $H_N := -\frac{1}{2N} \sum_{ij} \sgn(x_i)J_{ij} \sgn(x_j)$. We refer to this quantity as the decoded energy. Extending this framework, we incorporate a dummy source field into the generating functional; this field is conjugate to the decoded energy and serves to track its dynamical evolution.

Using DMFT \cite{Chow_2010, Grzetic_2014, Martin_1973} (see Appendix~\ref{appendx: DMFT_Calculation} for details), the mean field effective expression for the CIM is obtained in the spirit of Refs.~\cite{Sompolinsky_1981, Sompolinsky_1982}:
\begin{equation}
    \partial_t x(t)  =  - x^3(t) + ax(t) + J^2\xi^2 \int_{t_0}^{t} dt' G(t, t') x(t') + \phi(t).
    \label{eq:DMFT}
\end{equation}
Here, $\phi$ represents a colored noise with $\left\langle \phi_i(t)\phi_i(t') \right\rangle = 2T\delta(t-t') + J^2\xi^2{2}Q(t,t')$. The term $Q(t,t') := \langle x(t)x(t') \rangle$ is the correlation function, which characterizes the mean similarity between the soft spin states at times $t$ and $t'$. Meanwhile, $G(t,t') := \langle \partial x(t)/\partial h(t') \rangle$ is the response function, describing how the system at time $t$ responds to an infinitesimal perturbation applied at time $t'$. Due to causality, $G(t,t') = 0$ when $t' > t$. In comparison with Eq.~(\ref{eq:CIM}), this mapping replaces spatial couplings among soft spins with temporal correlations to their history. This construction extends naturally to other Ising machines such as those in Eq.~(\ref{eq:simCIM}), and (\ref{eq:digCIM}). For instance, adding a pre-processing clipping function $\varphi(x)=\max(-1,\min(1,x))$ to Eq.~(\ref{eq:DMFT}) yields clipCIM~\cite{Order-of-magnitude}, while removing the cubic term leads to simCIM~\cite{Tiunov:19, Zhou_2025} (i.e. Eq.~(\ref{eq:simCIM})).

We also obtain a closed DMFT expression for the decoded energy:
\begin{equation}
    \begin{aligned}
        E_{\text{dec}}(t) = -\int_0^{t}
        \left\langle \frac{\partial \sgn x(t)}{\partial h(t')} \right\rangle
        \left\langle \sgn x(t) x(t') \right\rangle dt' .
    \end{aligned}
    \label{eq:decoded_energy}
\end{equation}
\zs{This expression highlights the interplay between response and memory in the soft-spin dynamics. The decoded energy is evaluated from the final Ising variable \(\sgn x(t)\), whereas the response kernel describes its sensitivity to an infinitesimal field \(h(t')\) applied to the earlier continuous trajectory. The factor
    \(\langle \sgn x(t)x(t')\rangle\) is therefore a mixed memory correlator between the final decoded spin and the earlier soft-spin amplitude, not a decoded-spin autocorrelation.}

\begin{figure}[t]
    \centering
    \includegraphics[width=0.5\textwidth]{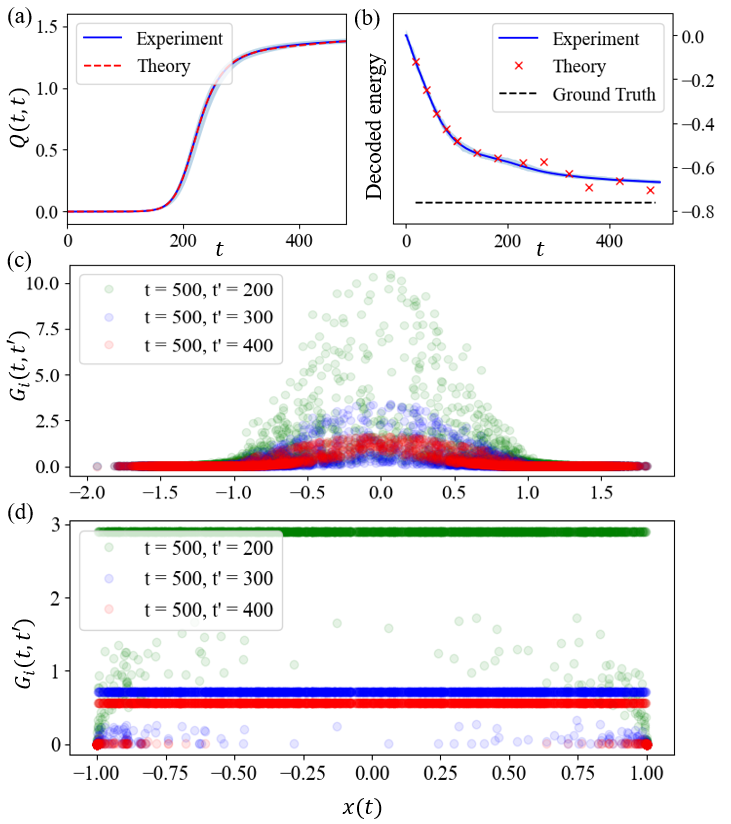}
    \caption{
        (a) Comparison of the diagonal terms of the correlation function $Q$ and the variance of soft spins measured in CIM experiments. (b) Comparison of the decoded energy obtained from Eq.~(\ref{eq:decoded_energy}) with experimental results from the CIM. The dashed line represents the ground state energy of the SK model in the thermodynamic limit. (c) Dependence of the response function $\partial x(t)/\partial h(t')$ on the value of soft spins $x(t)$.  (d) SimCIM’s dependence of the response function $\partial x(t)/\partial h(t')$  on the value of the soft spins $x(t)$.  \zs{Simulations from (a) -(c) were executed using the Euler-Maruyama method with a step size 0.02, $a = 0$, $T=10^{-5}$, $N=2000$, while \mw{in (d), we use} $a = -0.5$.}}
    \label{Fig.DMFT}
\end{figure}

\section{Time Dependence of the Decoded Energy}

We benchmark the theory by Monte Carlo simulations of both Eq.~(\ref{eq:CIM}) and Eq.~(\ref{eq:DMFT}). As shown in Fig.~\ref{Fig.DMFT}(a, b), the predictions for the diagonal of the correlation matrix $Q$ and the early-time decoded energy are in excellent agreement.
We note that the response function $\partial x(t)/\partial h(t')$ of individual nodes in CIM exhibit large variance in Fig.~\ref{Fig.DMFT}. Nevertheless, the Green's function $G(t, t')$ is well behaved after averaging $\partial x(t)/\partial h(t')$. The term $\langle \partial \sgn x(t)/\partial h(t') \rangle$ in Eq.~(\ref{eq:decoded_energy}) has the same behavior. Hence, the decoded energy can be obtained directly from Eq.~(\ref{eq:decoded_energy}). As shown in Eq.~(\ref{eq:decoded_energy}) in the large-$N$ limit, both the decay of the expected decoded energy toward its target and the expected time to reach it are independent of system size (see the blue line in Fig.~\ref{Fig.DMFT}(b) ) for a fixed parameter set, the expected hitting time for a target decoded energy $E_{\rm dec}$ is also size-independent, i.e., $t(E_{\rm dec})_{N} \equiv t(E_{\rm dec})$. In our numerical experiments, we confirmed that convergence of the dynamics can be achieved in $O(1)$ iterations. Each iteration involves matrix vector multiplication with $O(N^2)$ operations. Thus, the total number of operations required is $t(E_{\rm dec})N^2${\footnote{The non-monotonicity observed in the theoretical results at later evolutionary stages in Fig.~\ref{Fig.DMFT} is attributed to the finite sampling inherent in the Monte Carlo method. With increasing time, the response matrix grows quadratically while the near-zero density diminishes. This leads to increased variance in the estimator $\langle \partial \sgn x(t)/\partial h(t') \rangle$, resulting in statistical fluctuations.}.}

Even with a size-independent hitting time, the accessibility of a target decoded energy $E_{\rm dec}$ remains nontrivial. In particular, a practical issue is to determine the accessible target decoded energy window that an Ising machine can attain within a limited number of iterations. \zs{From Fig.~\ref{Fig.DMFT}(b), for both the theoretical and numerical results obtained with fixed $a=0, T=10^{-5}, N=2000$, a step size of 0.02, and 500 time steps, the lowest decoded energy reached remains well above the mean-field ground-state energy of the SK model.} A pragmatic approach is to benchmark by sampling a sufficient number of long-time runs, as we will show later. Below, we first provide theoretical insights that reveal fundamental mechanisms and performance barriers of Ising machines.

\section{Soft and hard spins}

An important element in the dynamics of the mean-field model is the history-dependent response function (or so called \zs{Onsager reaction term} \cite{Opper2001}) for each spin $x_i$, defined as the Green's function $G_i(t,t') := \partial x_i(t)/\partial h_i(t')$.  In Fig.~\ref{Fig.DMFT}(c)\footnote{Because $G_i(t,t')$ depends on the full trajectory, the figure appears as a scatter plot}, the heterogeneity of the responses increases with time. Figure~\ref{Fig.DMFT}(d) shows that the situation in simCIM (to be presented in a later section) is similar, except that bands of response functions appear due to the presence of bounds.

\zs{More significantly, Fig.~\ref{Fig.DMFT}(c) reveals a soft--hard dichotomy in the spin dynamics. We define a spin as soft when
    \[
        |x_i(t)|<x^\ast\mw{(t)},
    \]
    and as hard when
    \[
        |x_i(t)|\ge x^\ast\mw{(t)}.
    \]
For simCIM, \(x^\ast_t\equiv1\) naturally separates the near-zero soft-spin population from the large-amplitude hard-spin population. For CIM, where the amplitude distribution is broader and the separation is less sharply tied to a fixed amplitude scale, we define \(x^\ast_t\) through the contribution of large-amplitude spins to the response function. Specifically, \(x^\ast_t\) is \mw{related to a small prescribed tolerance $\epsilon$ by}
\begin{equation}
        \int_0^t\int_{|x\mw{(t)}|\ge x^\ast_t}
        \frac{\partial x(t)}{\partial h(t')}
        \,P(x\mw{(t)},t)\, {\rm d}x\mw{(t)}{\rm d}t'
        \le \epsilon.
\end{equation}
In this sense, hard spins are those whose contribution to the response is negligible, while soft spins carry most of the dynamical susceptibility.}

\zs{With this definition, small-amplitude spins remain close to the origin, exhibit large response, and are therefore susceptible to perturbations and zero-crossing events. Large-amplitude spins, by contrast, have much weaker response and are unlikely to change sign, behaving as dynamically hardened variables.}

\zs{This distinction is consistent \mw{with} and complementary \mw{to the previous} empirical ``swing/trapped'' observation of Ref.~\cite{Juntao2023}. In that work, swing and trapped nodes were numerically observed from the bifurcation dynamics of finite-size continuous Ising-machine trajectories: swing nodes remain near the origin and may oscillate between signs until later bifurcations determine them, whereas trapped nodes have already acquired large amplitudes and stable signs. Here, the soft--hard classification is formulated within the DMFT framework in the thermodynamic limit and is characterized by the amplitude distribution, response function, and zero-crossing activity. It therefore provides a mean-field dynamical explanation of the same phenomenology, while showing that the separation emerges self-consistently from the collective dynamics even at fixed gain.}

These observations lead us to interpret the dynamical process as follows: spins with large amplitudes will first stabilize, reducing the dynamics to a subspace spanned by the smaller-amplitude (soft) spins. These soft spins can then evolve rapidly and in parallel, consistent with the elevated response values observed in Fig.~\ref{Fig.DMFT}. Notably, this dichotomy can be even more pronounced in Ising machines that employ clipping mechanism $\varphi(x)$: once hard spins hit the boundary, they become effectively response-less and make zero contributions to the \zs{Onsager reaction term} (see Fig.~\ref{Fig.DMFT}(d) and Eq.~(\ref{eq:simDMFT})).

\section{Gap Opening}

In the above discussion, we argued that the final approach to the target decoded energy is dominated by the dynamics of soft spins within the reduced subspace. In the extreme limit where soft spins vanish altogether, the system freezes and the decoded energy stops evolving. At equilibrium, we refer to this phenomenon as “gap opening” as discussed in our previous work \cite{Zhou_2025}, or "hard phase" according to \cite{Yamamura2023}. Dynamically, this behavior can be further understood via Eq.~(\ref{eq:decoded_energy}). \zs{As derived in Appendix~\ref{appendx:Zero-crossing Rate}, one step zero-crossing rate $\mu(t)$ can be approximately expressed as} 
\mw{
\begin{equation}
    \sqrt{\Delta t}\mu(t)= \frac{\sqrt{T}}{\sqrt{\pi}}P(0).
\label{eq:effective_gap}
\end{equation}
The scaling of $\sqrt{\Delta t}$ appears because, in the presence of a white-noise component, the measured zero-crossing rate at finite time resolution scales as $\mu(t) \propto 1/\sqrt{\Delta t}$. Multiplying by $\sqrt{\Delta t}$ removes this trivial time-resolution dependence. Note that the right-hand side solely depends on the intrinsic characterization of the model (such as  $T$ and $P(0)$), and dependence on simulation parameters (such as $\Delta t$) is decoupled. Thus, $\sqrt{\Delta t} \mu(t)$ becomes a more intrinsic quantity for comparing contour levels. Hereafter, it will be referred to as the {\it effective gap}.}

When a gap opens in the distribution at a specific instant $t'$, the spin density at $x(t') = 0$ vanishes, rendering further spin flips impossible. Consequently, the response term $\langle \partial \sgn x(t)/\partial h(t') \rangle = 2 \langle \delta(x(t)) \partial x(t)/\partial h(t') \rangle$ also vanishes for all subsequent times $t > t'$, as the trajectory $x(t)$ no longer contributes to the Dirac delta distribution at 0. Under these conditions, the decoded energy ceases to decrease. Therefore, maintaining a finite density of near-zero spins throughout the evolution is crucial to avoid dynamical stagnation and ensure continuous energy reduction.

Equilibrium analysis shows that, in the thermodynamic limit, a true gap (vanishing zero‑crossing rate) opens only in the noiseless limit, $\beta \to \infty$. In practice, however, CIM dynamics always operate with a small but nonzero noise level, either via deliberate injection or imposed by hardware constraints~\cite{PhysRevA.88.063853}. Under these conditions, any gapped spin‑amplitude distribution is smoothed: noise continually replenishes the zero‑crossing rate, which therefore never strictly vanishes. As a result, a rigorously defined gap does not arise in dynamics, and the updating of Ising states never fully halts.

Even so, for finite systems the evolution can become too slow to matter, i.e., a practical infinity for $t(E_{\rm dec})$ for desired values of $E_{\rm dec}$. While updates persist in principle, the density of near-zero spins can drop to negligible levels, rendering the dynamics effectively frozen. We use the ensemble zero-crossing rate \mw{$\sqrt{\Delta t}\mu(t)$ as the primary indicator of the effective gap}. {\it Thus, a sufficiently small $\mu(t)$ signals the onset of the effective gap, where meaningful evolution has essentially ceased, and provides a sufficient condition for the practical inaccessibility of a target energy $E_{\rm dec}$.}

\begin{figure}[ht]
    \centering
    \includegraphics[width=0.5\textwidth]{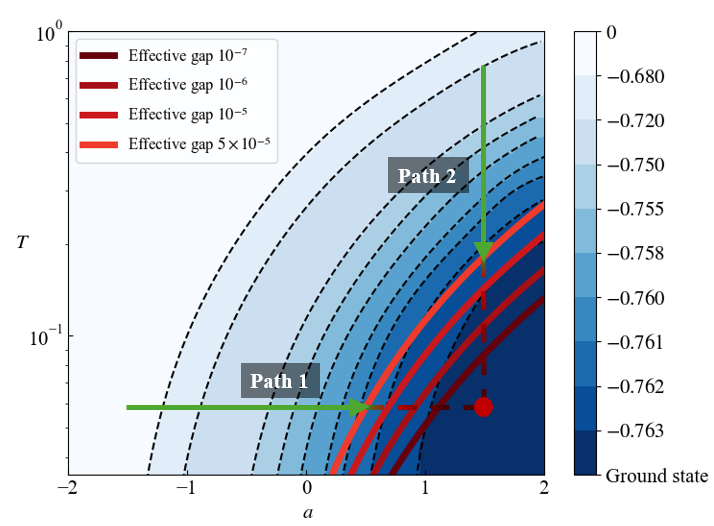}
    \caption{The decoded energy contours and effective gap lines of the CIM. Path 1 represents the conventional strategy of adjusting the parameter $a$, while Path 2 represents annealing solely by temperature. \zs{Both paths aim at the same termination point (the red dot) by schedule but are prematurely halted by the effective gap. }
        Notably, path 2 reaches a lower decoded energy than path 1 when they are halted at the same effective gap contour.}
    \label{Fig.zero_density}
\end{figure}

\mw{To connect the effective gap with quantities accessible in simulations, we refer to the time scale $1/(N\mu(t))$ as the time budget $\tau_{\rm budget}$, which is the time scale taken by the simulation model to flip one spin \cite{Rice_1944}. When the time budget becomes of the order 1 or longer, the dynamics is effectively frozen. The time budget is related to the effective gap via
\begin{equation}
	\tau_{\rm budget} = \frac{\sqrt{\Delta t}}{N[\sqrt{\Delta t}\mu(t)]}.
\end{equation}
Figure~\ref{Fig.zero_density} shows the contours of the effective gaps for selected small values in the $(a, T)$ space.}
\zs{These contours identify regions where spin flips are statistically unlikely within the available computational budget. The plot is obtained for CIM at quasi-equilibrium, based on our theoretical calculations.\footnote{We utilized the one-step replica-symmetry-breaking (1RSB) formula derived in our previous work to generate the curves~\cite{Zhou_2025}. The theoretical values may exhibit a small bias relative to the true full-RSB results, but they are sufficient to capture the overall trend. More precise but qualitatively similar results can be obtained by running Monte Carlo simulations according to Eq.~\mw{(\ref{eq:DMFT})} for sufficiently long durations and large samples.} We numerically verify this trend of \mw{the} effective gap on Fig.~\ref{Fig.zero_density}.}
\mw{Specifically, using the values of $P(0)$ derived in the 1RSB approximation, we expect that in our simulation setting, the effective gap at which the dynamics is frozen lies in the range between $10^{-6}$ and $10^{-5}$. This agrees well with simulations searching for the freezing points of annealing in Fig.~\ref{Fig:MC} of Appendix~\ref{appendix:num}.}

\mw{The contour plots elucidate }\zs{the discrepancy between zero-temperature equilibrium theory and experimental observations, while providing practical guidance for parameter scheduling in Ising machines. Our analysis particularly demonstrates that, for CIMs with a fixed deeply buried target point, slow temperature annealing constitutes a better strategy.}

A common CIM implementation initializes the parameter $a$ with a small value where the energy landscape is sufficiently simple to facilitate finding the global minimum at the beginning. The parameter $a$ is subsequently increased gradually~\cite{Mohseni2022}, while maintaining constant low noise. It is commonly assumed that sufficiently slow adjustment of $a$ enables the Ising machine to preserve its ground state despite increasing landscape complexity, consistent with adiabatic evolution principles~\cite{Leleu_2019, PhysRevA.88.063853}.

\begin{figure}[htb]
    \centering
    \includegraphics[width=0.5\textwidth]{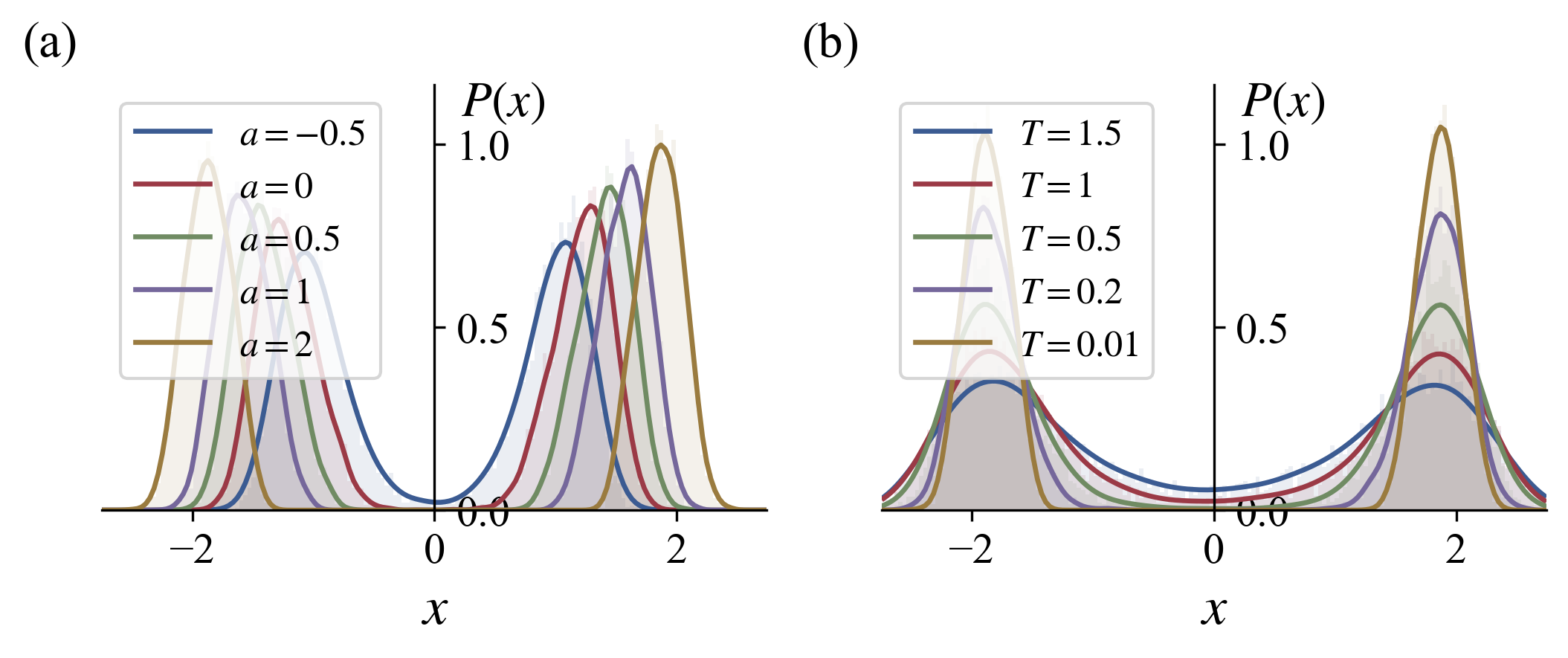}
    \caption{An illustration of the distributional evolution of CIM. Panel (a) fixes \mw{$T = 0.01$} and anneals \mw{$a$}; panel (b) fixes \mw{$a = 2$} and anneals \mw{$T$.}}
    \label{Fig.CIM_dis}
\end{figure}

However, equilibrium analysis reveals that the ground state of the CIM can only be reached at significantly large $a$ and $T = 0$~\cite{Juntao2023}. As demonstrated by~\cite{Yamamura2023}, in the noiseless case, the parameter $a$ schedule should terminate at $-0.45$ for the SK model in CIMs, corresponding to the true gap, where all near-zero soft spins vanish. While noise enables Ising machines to progress beyond this threshold, the effective gap nevertheless impedes further reduction in decoded energy. Specifically, once the system traverses the effective gap boundary, the timescale required to maintain the ground state or target energy $E_{\rm dec}$ becomes computationally prohibitive, scaling at least as $O(1/N\mu)$. \zs{Beyond this critical point, further increasing parameter $a$ yields no performance enhancement in practice and instead leads to a longer convergence time.}

\zs{The gap-formation process shown in Fig.~\ref{Fig.CIM_dis}, together with the single-run trajectories in Fig.~\ref{Fig.single_experiment} of Appendix~\ref{appendx: Experiments_Setting}, illustrates this difference. \mw{Suppose the target decoded energy is $10^{-2}$ above the ground state energy.} For gain annealing, as shown in \mw{Fig.~\ref{Fig.CIM_dis}(a) and  Fig.~\ref{Fig.single_experiment}(b)}, the gap opens \mw{at around $a = 0$. Figure~\ref{Fig.single_experiment}(a) shows that the CIM has not reached the target decoded energy. By} contrast, for temperature annealing, as shown in \mw{ Fig.~\ref{Fig.CIM_dis}(b) and  Fig.~\ref{Fig.single_experiment}(b)}, the spin density at \mw{$x=0$} remains nonzero until the CIM reaches the target decoded energy, preserving a finite probability of further sign changes and equilibration.}

With the help of the effective gap contours in the phase space, we are able to design a more effective annealing schedule that involves not only adjusting the parameter $a$ but also the temperature. From the previous analysis, one can overlay the effective gap contours with theoretically achievable decoded energy contours. As shown in Fig.~\ref{Fig.zero_density}, we observe that the energy contours are more slanted compared to the effective gap contours. \zs{This indicates that traditional strategies involving horizontal adjustments of parameters (modifying $a$ alone) at a low $T$ are more likely to be halted early by the effective gap. In contrast, vertical strategies (temperature annealing) are affected at a later stage and offer a more reliable path to optimization \footnote{This also suggests choosing the gain at the point where the decoded energy contour is tangent to the effective gap contour.}.}

\begin{figure}[htb]
    \centering
    \includegraphics[width=0.48\textwidth]{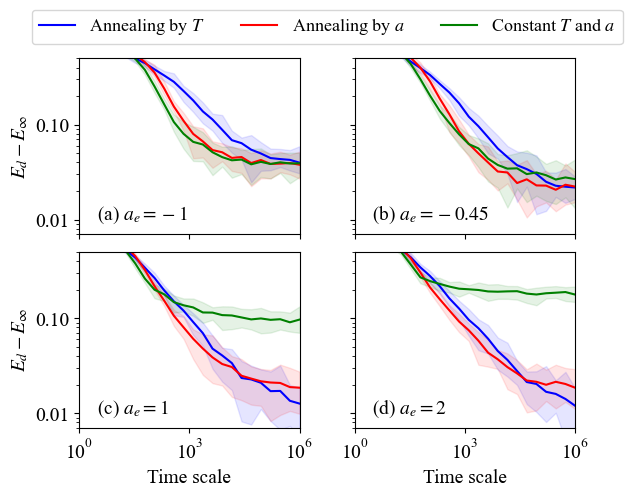}
    \caption{Comparison of decoded energy profiles obtained via different parameter-tuning strategies across varying time scales.  Data points represent the mean value of the decoded energy from the final step across 20 independent runs with the same annealing protocol for SK models of $N= 2000$. Detailed parameter settings can be found in Appendix~\ref{appendx: Experiments_Setting}. The dashed line represents the ground truth. $a_e$ marks the terminal point of $a$. Panel (a) corresponds to the gapless regime, (b) aligns with the gap opening at zero temperature, and (c) and (d) represent gap-opened regime.  Shaded areas indicate the range of maximum and minimum values observed
        \label{Fig.numerical_result}. \zs{The ground state $E_\infty$ is set to be $-0.7632$}.}
\end{figure}

This result is consistent with numerical experiments. As shown in Fig.~\ref{Fig.numerical_result} for the SK model, we compare three parameter-tuning strategies: (i) increasing $a$ to $a_e$ at a fixed, low temperature (Path 1 in Fig.~\ref{Fig.zero_density}), (ii) annealing the temperature while keeping $a_e$ high and constant (Path 2 in Fig.~\ref{Fig.zero_density}), and (iii) maintaining both $T$ and $a_e$ as constants. Each strategy was independently tested with varying total run times, as
represented on the x-axis. Longer time scales yield smoother parameter adjustments and improved performance.

In Fig.~\ref{Fig.numerical_result}(a) and (b), the terminal points lie outside the region enclosed by the effective gap contour of  $O(\frac{1}{N})$, avoiding early termination by the contour. Consequently, with sufficiently long run times, adjusting either $T$ or $a$ achieves comparable performance. In contrast, in Fig.~\ref{Fig.numerical_result}(c) and (d), early termination by the effective gap causes $T$ annealing to outperform $a$ tuning on longer time scales. More details about experiment settings and results are provided in the Appendix \ref{appendx: Experiments_Setting}.

\section{Exposed versus Hidden Decoded Energy}

As described in the previous section, if the state of the target decoded energy is {\it hidden} in the region enclosed by the effective gap contour in the space of $a$ and $T$, then the annealing process, either by annealing the gain or temperature or both, cannot access the target state, and we observe that for CIM, temperature annealing ends up at a lower decoded energy on the effective gap contour. On the other hand, if the state with the target decoded energy is {\it exposed} by the effective gap contour, i.e., located at the boundary or outside the hidden region, then scheduling $a$ can achieve comparable or even better performance than scheduling $T$.

We demonstrate such a behavior in simCIM~\cite{Tiunov:19}, which has been implemented on digitized hardware. Its dynamics is governed by
\begin{equation}
    \frac{dx_i}{dt} = a\varphi (x_i) + \sum_j J_{ij}\varphi(x_j),
    \label{eq:simCIM}
\end{equation}
where the clipping function is defined as $\varphi(x) =x$ for $|x|\le 1$ and $\sgn x$ otherwise, and the cubic saturation term is avoided in the dynamics. The DMFT expression of simCIM can be derived as the same manner, it reads:
\begin{equation}
    \begin{aligned}
         & \partial_t x(t)  =  ax(t) + J^2\xi^2 \int_{t_0}^{t} dt' G(t, t') x(t') + \phi(t); \\ &x(t) = \varphi(x(t)). \label{eq:simDMFT}
    \end{aligned}
\end{equation}
The decoded energy expression is also consistent with Eq.~(\ref{eq:decoded_energy}), while the soft–hard spin classification and the response function distribution are shown in Fig.~\ref{Fig.DMFT}(d).

\begin{figure}[htb]
    \centering
    \includegraphics[width=0.45\textwidth]{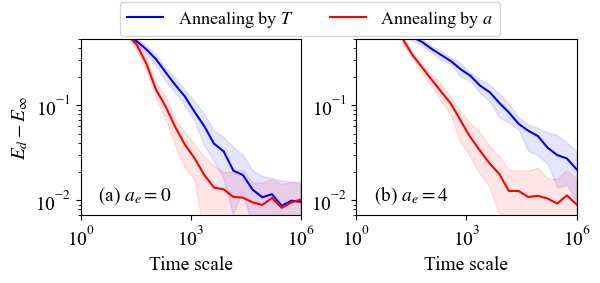}
    \caption{Comparison of decoded energy profiles obtained via different parameter-tuning strategies across varying time scales for simCIM. Each data point consists of 20 independent trials conducted under different total running time scales, as represented on the $x$-axis.}
    \label{Fig:experiment_simCIM}
\end{figure}

In simCIM, the optimal target decoded energy is located at the gapless-binary transition point $(a, T) = (0, 0)$~\cite{Zhou_2025}, implying that it lies exactly on the effective gap contour with $P(0) = 0$. Envisaging the adoption of an effective gap contour with $P(0)$ approaching 0 in numerical experiments, it is possible that the annealed states can get arbitrarily close to the optimal point. In such cases, appropriately annealing by either temperature or gain can lead to decoded energies approaching optimum, as shown in Fig.~\ref{Fig:experiment_simCIM}(a).

Apparently, this picture of an exposed optimal state is inconsistent with the gap-opening scenario presented in the previous section, which is common in many Ising machines. This is because normally the spins of the optimal state \zs{have} a binary distribution separated by a wide gap, implying that $P(0)$ should have vanished in the vicinity of the optimal state in the space of $a$ and $T$.


\begin{figure}[htb]
    \centering
    \includegraphics[width=0.45\textwidth]{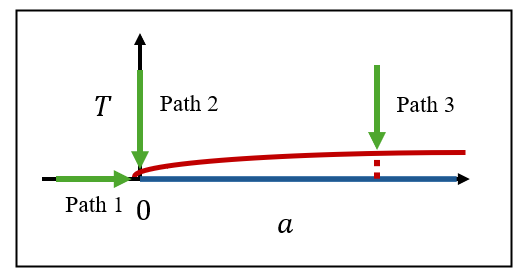}
    \caption{Schematic plot of the decoded energy contour at the ground state (blue line) and an effective gap contour (red line) of simCIM. The effective gap contour is closely aligned with the ground-state contour near the origin. }
    \label{Fig.zero_density_sim}
\end{figure}

How can the optimal state in simCIM be exposed by the effective gap? Remarkably, gap formation in simCIM goes through an alternative scenario, in which the spin distribution in the neighborhood of $x = 0$ is continuously reduced when $a$ increases, instead of disappearing discontinuously. From equilibrium analysis~\cite{Zhou_2025}, we establish that at zero temperature, near-zero soft spins in the simCIM system vanish only when the ground state becomes accessible, corresponding to $(a, T) = (0, 0)$ \footnote{\mw{The result $(a, T) = (0, 0)$} includes the full-RSB correction. See 
\cite{Zhou_2025}}. Prior to this point, a finite zero-crossing rate persists even in the absence of noise. We conceptually illustrate the relationship between the decoded energy contours and the effective gap contour for simCIM in Fig.~\ref{Fig.zero_density_sim}.  Consequently, near $(a, T) = (0, 0)$, the effective gap boundary closely coincides with the lowest decoded energy contour, as illustrated at the terminal of Path 1 and Path 2 in Fig.~\ref{Fig.zero_density_sim}. This alignment implies that slow annealing of $a$ yields comparable performance to temperature annealing when terminating at $(a, T) = (0, 0)$, as evidenced by Fig.~\ref{Fig:experiment_simCIM}(a). However, for termination at $a > 0$ and $T = 0$, continuous annealing of $a$ provides no performance enhancement. Conversely, temperature annealing exhibits reduced efficacy due to the effective gap emergence (see Path 3 and the blue curve in Fig.~\ref{Fig:experiment_simCIM}(b)).

\zs{Although the DMFT derivation above is exact only for the mean-field SK structure, we also test whether the same qualitative effective-gap mechanism remain useful \mw{empirically} beyond this setting. Numerical evidence on the Gset benchmark is presented in Appendix~\ref{appendix:gset}.}

\section{Scaling Function of Decoded Energy}

\begin{figure}[htb]
    \centering
    \includegraphics[width=0.5\textwidth]{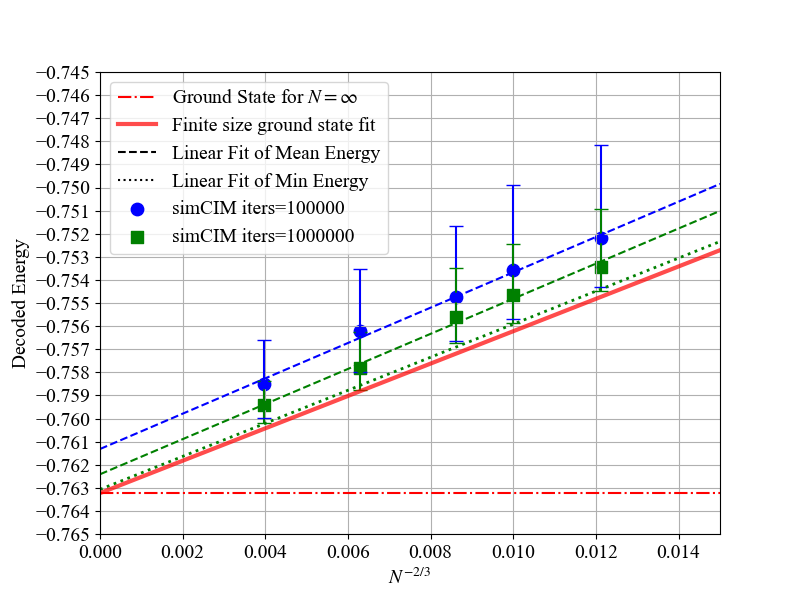}
    \caption{Energies obtained by simCIM using the Euler–Maruyama method (step size 0.1) with 100,000 and 1,000,000 iterations. For each system size, N = 750, 1000, 1250, 2000, and 4000, we evaluate 200 independent SK instances and perform 30 runs per instance. Data points report the decoded energy averaged over the 30 runs and then averaged across instances. Error bars show, for each instance, the mean of the minimum and maximum decoded energies observed over the 30 runs. The red line denotes the mean ground-state energy with finite-size corrections reported in Ref.~\cite{Boettcher2010}. The green dotted line indicates the fit for the expected minimum value after $10^6$ iterations, yielding an extrapolated infinite-size value of $-0.76307 \pm 0.00018$ }
    \label{Fig.sim_scale}
\end{figure}

We have discussed the planning of optimal pathways for annealing in Ising machines. The remaining issue is how the resultant decoded energy depends on time. Remarkably, for a family of models, we note that the finite-time decoded energy of a model approaches its asymptotic value at the same rate as the family of models in the large $N$ limit. This implies that knowing the convergence behavior of the large $N$ models is useful for estimating the convergence behavior of the models of arbitrary size.

To see this, we note that Eq.~(\ref{eq:decoded_energy}), derived from DMFT, implies that the decoded energy decay is independent of system size. While strictly valid only in the thermodynamic limit ($N\rightarrow\infty$), we observe that the energy density evolution is largely insensitive to finite-size effects for large $N$. This justifies the approximation:
\begin{equation}
    E_{\rm dec}(t, N) - E_{\rm dec}(\infty, N)
    \approx \lim_{N\rightarrow\infty}\left(E_{\rm dec}(t, N) - E_{\rm dec}(\infty, N) \right),
\end{equation}
where $E_{\rm dec}(t, N)$ denotes the decoded energy obtained by the Ising machine at time $t$ for a system of size $N$. Essentially, the relaxation of the decoded energy toward equilibrium in a finite-size system approximates the behavior of the system in the thermodynamic limit. Consequently, in conjunction with the well-characterized scaling behavior of the SK model, one can investigate the solution quality at finite time for arbitrary $N$. The finite-size scaling of the equilibrium ground-state energy for the SK model follows an $N^{-2/3}$ power law, as derived theoretically~\cite{Aspelmeier_2007} and numerically~\cite{Boettcher2004, Kim2007, Palassini_2008}:
\begin{equation}
    E_{\rm dec}(\infty, N) \approx E_\infty  + A N^{-2/3},\label{eq:finite_scaling}
\end{equation}
where $E_\infty = -0.7632$ represents the ground-state energy of the infinite-size SK model and $A = 0.701$ \cite{Palassini_2008, Boettcher2010}.

Here, we apply simCIM, generally believed to outperform CIM \cite{Bohm2019-pi, Zhou_2025}, by running 30 trials for each of 100 relaxations of SK instances at various system sizes, using path 2 as described in Fig.~\ref{Fig.zero_density_sim}.

\begin{figure}[h]
    \centering
    \includegraphics[width=0.5\textwidth]{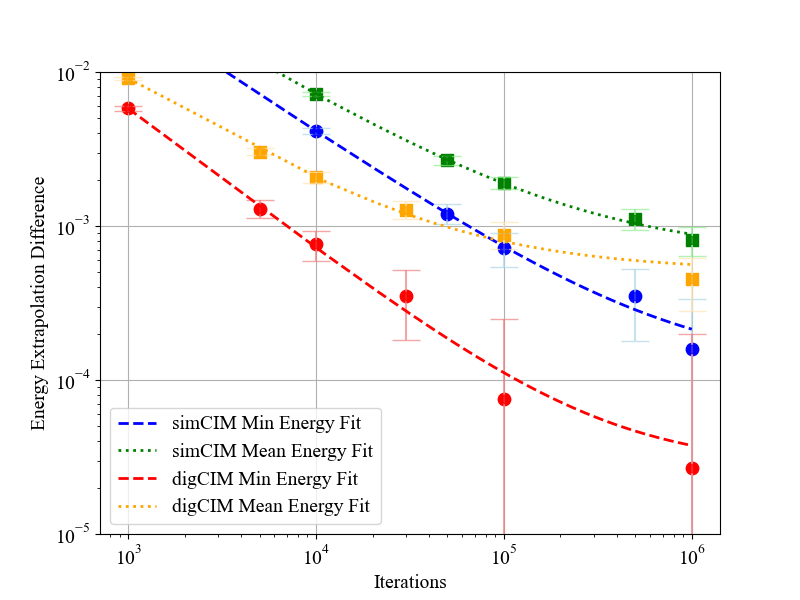}
    \caption{The decoded energy \( \lim_{N\to\infty} E_{\text{dec}}(t, N) \equiv E_{\text{dec}}(t, \infty) \), extrapolated to the infinite-size limit (\( N\to\infty \)) at various iteration steps for simCIM and digCIM. Experimental parameters are consistent with Figs.~\ref{Fig.sim_scale} and \ref{Fig.dig_scale}. Squares indicate the mean energy, extrapolated from the average of 30 trials across 200 independent SK instances. Dots represent the minimum energy, extrapolated from the best result obtained in 30 trials. Error bars indicate the standard error of the extrapolated values. Solid lines indicate fits to Eq.~\ref{eq:power_law} for each dataset. Data are offset by -0.76323\cite{Boettcher2004, Boettcher2010}, a slightly lower estimate of the SK ground-state energy, rather than by the individual fitted bias parameters. The $y$-axis represents the deviation from this reference value on a logarithmic scale.}
    \label{Fig.dig_sim_iter}
\end{figure}

As shown in Fig.~\ref{Fig.sim_scale}, the decoded energies for given values of $t$ have linear relationships with $N^{-2/3}$, sharing the same slope $A$ with the finite-size scaling of the ground state energy. Extrapolating the linear relations to the intercept, we obtain the thermodynamic limit of the decoded energies of finite times. {We observed that the average and minimum values of $E_{\rm dec}(t, \infty) $ versus $t$ can be approximated as power laws.} In particular,
\begin{equation}
    E_{\rm dec}{(t, \infty)} \approx E_\infty  + B t^{-\xi}.\label{eq:power_law}
\end{equation}
As shown in Fig.~\ref{Fig.dig_sim_iter}, $E_{\rm dec}(t, \infty)$ {has a very good power-law fit with $t$ after subtracting} the bias term $E_\infty$. For average values of $E_{\rm dec}(t, \infty) $, we obtain $E_\infty = -0.7626 \pm 0.0003$, $B = 4.657$, and $\xi = 0.713$,
whereas for minimum values of $E_{\rm dec}(t, \infty)$, we get $E_ \infty = -0.7632 \pm 0.0003$, $B =  4.656$, and $\xi = 0.763$. Thus, Eqs. (\ref{eq:finite_scaling} - \ref{eq:power_law}) allow us to estimate the decoded energy
at finite time for arbitrary $N$\zs{. Equivalently, if $\epsilon=E_{\rm dec}(t,\infty)-E_\infty$ denotes the target energy-density error, Eq.~(\ref{eq:power_law}) gives $t(\epsilon)\sim \epsilon^{-1/\xi}$. The corresponding SK computational cost is therefore $\mathrm{Cost}(N,\epsilon)=C_{\mathrm{sim/dig}}(\epsilon)N^2\sim N^2\epsilon^{-1/\xi}$, rather than a uniform $O(N^2)$ cost as $\epsilon\to0$; the $O(N^2)$ statement applies at fixed target accuracy.}

Finally, applying the scaling relations of the dynamical cases enables us to compute the
ground state energy of the Ising model at the thermodynamic limit with accuracy
matching state-of-the-art measurements \cite{Boettcher2004, Kim2007, Palassini_2008,
    Boettcher2010}. This is achieved by directly reading the value of $E_ \infty$ from Eq. \ref{eq:power_law} for
fitting the minimum values of $E_{\rm dec}(t, \infty)$, as finding
the minimal decoded energy  in practice is a more common decoding method than finding the mean value.

The value of $E_ \infty = -0.76317 \pm 0.00029$ is close to the numerical estimate of the
ground-state energy of -0.7632 \cite{Boettcher2004, Kim2007, Palassini_2008,
    Boettcher2010} and the theoretical estimate of -0.76317 \cite{oppermann_2007}. It is
also numerically consistent with the results of Ref.~\cite{Alaoui_2020}, where a message-passing algorithm is claimed to achieve near-optimal solutions arbitrarily close to the ground state in $O(N^2)$ time.

An even more accurate estimate of the ground-state energy is provided by digCIM
\cite{Zhou_2025} with shorter convergence time and higher success probability. The
estimate by digCIM is $-0.76317 \pm 0.00015$ (see Appendix \ref{appendix:DigCIM Results}). Besides our work, several experimental variants of Ising machines report superior performance, including digCIM \cite{Zhou_2025}, dSB \cite{Goto2021}, and CAC \cite{Leleu_2019}. \zs{Appendix~\ref{appendix:advanced_solvers} provides a phenomenological comparison with dSB and CAC. In these examples, $P(0)$ remains finite while the decoded energy decreases, suggesting that these mechanisms may help sustain soft-spin activity and postpone dynamical hardening.}

\section{Conclusion}
In summary, we analyzed the behaviors of dynamical optimization solvers by deriving their dynamical mean-field descriptions. To obtain a theoretically meaningful description for finite-size systems, we introduced the concept of an effective gap, which can impede further energy reduction. We proposed using the density of near-zero soft spins as an indicator for the onset of this effective gap. By comparing decoded-energy contours with the emergence of the gap, we designed an optimal parameter-scheduling strategy for Ising machines. Notably, in contrast to the conventional approach of annealing the gain, the optimal schedule for CIM
annealing is through annealing the temperature. We also showed that this framework naturally generalizes to other platforms. On the other hand, our analysis of simCIM indicates that if the target decoded energy is exposed by the effective gap contour, annealing the parameter a is also effective. Across all cases, our theoretical predictions are in strong agreement with numerical experiments. {We mention in passing that temperature annealing is also more advantageous than gain annealing in digCIM. We have shown that since system states in the regime of negative gain are isomorphic with each other in digCIM \cite{Zhou_2025}, annealing the gain offers little help for convergence to the optimal solution, in contrast with temperature annealing.}

Building on this, we argued that Ising machines can reach near-optimal energies within time complexity $t(E_{\rm dec}) N^2$, \zs{so their fixed-accuracy $N$-scaling has the same SK $N^2$ form as rigorously analyzed message-passing methods~\cite{Alaoui_2021}, while the accuracy dependence is set empirically by $t(\epsilon)\sim \epsilon^{-1/\xi}$ rather than by a proven matching bound.} We further identified two key ingredients underlying this performance: the fixing of hard spins and simultaneous updating soft spins. In addition, we clarified the pivotal role of the zero-crossing rate of soft spins in driving improvements of the decoded energy.

This study focused on gradient-descent dynamics. In machines using Hamiltonian
dynamics with momentum, we expect similar asymptotic time-complexity advantages but with smaller constant factors, and extra parameter scheduling; exploring this regime is an interesting direction for future work.

\begin{acknowledgments}
This work was supported in part by Huawei through a CSTT project. \zs{During the finalization of this manuscript, we became aware of related parallel work on the geometry and dynamics of gain-annealed CIMs in SK-related models with planted solutions, which also employs DMFT \cite{ghimenti2025geometry}. A detailed comparison is provided in Appendix~\ref{appendix:related_work}.}
\end{acknowledgments}

\bibliography{ref.bib}
\appendix

\section{DMFT Calculation}
\label{appendx: DMFT_Calculation}

\zs{We first reformulate Eq.~\ref{eq:CIM} as a Martin--Siggia--Rose--Janssen--De Dominicis (MSRJD) functional integral
    \cite{Martin_1973,Janssen_1976,Dominicis_1976,Dominicis_1978}. The resulting action closely follows the dynamical formulation of Sompolinsky and Zippelius
    \cite{Sompolinsky_1981,Sompolinsky_1982}, with the additional source term \(l_e(t)\) coupled to the decoded Ising energy. As shown below, this source field allows us to generate the decoded-energy observable within the DMFT formalism and introduces the corresponding additional order parameters:
    \begin{widetext}
        \begin{equation}
            \begin{aligned}
                Z[t|t_0]
                = & \int Dx\,D\hat{x}\,
                \prod_i \exp \Bigg\{
                \int dt \Big[
                    l_i(t)x_i(t)
                    + l_e(t)\sum_{i<j}J_{ij}\sgn x_i(t)\sgn x_j(t)
                \\
                  & \quad
                    + i\hat{x}_i(t)
                    \left(
                    -\partial_t x_i(t)
                    -x_i^3(t)
                    +a x_i(t)
                    +\xi\sum_j J_{ij}x_j(t)
                    \right)
                    -T\hat{x}_i^2(t)
                    -\frac{1}{2}\left(3x_i^2(t)-a\right)
                    \Big]
                \Bigg\}.
            \end{aligned}
        \end{equation}
    \end{widetext}
    Here \(\hat{x}_i(t)\) is the MSRJD auxiliary response field. It enforces the stochastic equation of motion in the functional-integral representation and, physically, correlations involving \(\hat{x}_i(t)\) generate linear response functions of the trajectory \mw{responding} to infinitesimal perturbations. The fields \(l_i(t)\) and \(l_e(t)\) are source fields conjugate to \(x_i(t)\) and to the decoded Ising energy
    \(\sum_{i<j}J_{ij}\sgn x_i(t)\sgn x_j(t)\), respectively. In particular, \(l_e(t)\) is introduced as a generating field that enables the extraction of \(E_{\rm dec}(t)\) from the dynamical action. Finally, the term
    \[
        \exp\left[-\frac{1}{2}\int dt\,\left(3x_i^2(t)-a\right)\right]
    \]
    is the Jacobian associated with transforming the stochastic path integral from noise variables to dynamical variables in the Stratonovich convention
    \cite{Dominicis_1978,Bausch_1976}.}

By integrating over the coupling matrix and leveraging the properties of the SK model, we reorganize the terms to obtain the disorder average:
\begin{widetext}
    \begin{equation}
        \begin{aligned}
             & \mathbb{E}_{J_{ij}}(Z[t|t_0]) = \int Dx D\hat{x} \prod_{i} \exp ( S^0_i ) \times \prod_{i<j} \exp \left(\frac{J^2}{2N} \Sigma_{ij}^2 \right),                                 \\
             & S_i^0=\int dt\left\{  l_i(t) x_i(t)  + i\hat{x_i}(t)\left[-\partial_t x_i(t) - x_i^3(t) + ax_i(t)  \right] -\frac{1}{2}\left[3x_i^2(t) - a \right]  -T\hat{x_i}^2(t)\right\}, \\
             & \Sigma_{ij}= \int dt [\xi i\hat{x_i}(t) x_j(t) + \xi i\hat{x_j}(t) x_i(t)
                    + l_e(t)\sgn x_i(t)\sgn x_j(t)].
        \end{aligned}
    \end{equation}
\end{widetext}
The interaction terms can be decoupled using the Hubbard-Stratonovich transformation. After reorganization, we obtain
\begin{widetext}
    \begin{equation}
        \begin{aligned}
             & \prod_{i<j} \exp\left(\frac{J^2}{2N}\Sigma_{ij}^2\right)
            = C\left(\prod_{k=1}^7 \int dq_k(t_1, t_2)\right) \exp\left\{
            -NJ^2\int dt_1 \int dt_2
            \left[\frac{\xi^2}{2} q_1 q_2 +\frac{\xi}{2} q_3 a_4 +q_5 q_6 +\frac{1}{4}q_7^2\right.\right. \\
             & + J^2\sum_i\left(
            \frac{\xi^2}{2} q_1 i\hat x_i(t_1) i\hat x_i(t_2)
            + \frac{\xi^2}{2} q_2 x_i(t_1) x_i(t_2)
            + \frac{\xi^2}{2} q_3 i\hat x_i(t_1) x_i(t_2)
            + \frac{\xi^2}{2} q_4 x_i(t_1) i\hat x_i(t_2)\right.                                          \\
             & \left.\left.\left. + l_e(t_1) q_5 {\rm sgn} x_1(t_1) i\hat x_i(t_2)
            + q_6 {\rm sgn} x_i(t_1) x_i(t_2)
            + \frac{1}{2} \sqrt{l_e(t_1) l_e(t_2)} q_7 {\rm sgn}x_i(t_1) {\rm sgn}x_i(t_2)\right)\right]\right\}.
        \end{aligned}
    \end{equation}
\end{widetext}
where $C$ is a constant generated by the transformations and does not affect subsequent calculations. The auxiliary variables $q_k$ are introduced to decouple quadratic terms such as $i\hat{x}(t), x_j(t)$ and can be validated via the saddle-point conditions, e.g., $\partial \mathbb{E}(Z)/\partial q_k = 0$. Here, $D$ denotes the Gaussian measure. After enforcing causality and integrating out the auxiliary variables, the only order parameters with direct physical meaning are the correlation functions ($q_1$ and $q_5$) and the response functions ($q_3$, $q_4$ and $q_6$). Since we can always set the auxiliary variables $l_e$ and $\hat{x}$ to zero, the remaining auxiliary order parameters can likewise be taken to vanish. From the saddle-point equations, their explicit forms are given by
\begin{equation}
    \begin{aligned}
         & q_1(t_1, t_2) = \left\langle x(t_1)x(t_2)\right\rangle;                                 \\
         & q_3(t_1, t_2) = \left\langle x(t_1) i\hat{x}(t_2)\right\rangle \text{ for\ } t_1 > t_2; \\
         & q_4(t_1, t_2) = \langle i\hat x(t_1) x(t_2)\rangle \text{ for\ } t_1 < t_2;             \\
         & q_5(t_1, t_2) = \langle {\rm sgn} x(t_1) x(t_2)\rangle;                                 \\
         & q_6(t_1, t_2) = l_e(t_1) \left\langle \sgn x(t_1)i\hat{x}(t_2)\right\rangle
        \text{ for\ } t_1 > t_2.
    \end{aligned}
\end{equation}
The first three terms can be used to recover the mean-field dynamics, Eq.~(\ref{eq:DMFT}), as derived by Sompolinsky \cite{Sompolinsky_1982}. Meanwhile, the latter order parameter provides the dynamical mean-field expression for the decoded energy, Eq.~(\ref{eq:decoded_energy}), as follows:
\begin{equation}
    \begin{aligned}
         & E_{dec}(t) = \frac{\partial \mathbb{E}_{J_{ij}}(Z[t|t_0])}{\partial l_e(t)}                                     \\
         & = -J^2 \int_0^{t} dt' \left\langle\sgn x(t) x(t')\right\rangle\left\langle \sgn x(t)i\hat{x}(t')\right\rangle . \\
    \end{aligned}
\end{equation}

\section{The Zero-crossing Rate}
\label{appendx:Zero-crossing Rate}

Consider the dynamical equation Eq.~(\ref{eq:DMFT}) when the Ising machine approaches equilibrium, which becomes
\begin{equation}
    \frac{dx}{dt} \approx (a + \chi) x + \phi(t),
\end{equation}
where the susceptibility $\chi$ is the steady-state value of the integral $\int dt' G(t, t')$. In discrete-step dynamics, the dynamics becomes
\begin{equation}
    x(t + \Delta t) = x(t) + \Delta t \mw{[}a_{\rm eff} x(t) + \sqrt{\langle\phi(t)^2\rangle} z(t)\mw{]},
\end{equation}
where we further denote the effective gain $a_{\rm eff} = a + \chi$, and $z(t)$ is a Gaussian variable of mean 0 and unit variance. Suppose the spin at time $t$ is $x(t) = -\epsilon$. Then the condition that $x(t + \Delta t)$ changes sign is
\begin{equation}
    -\epsilon + \Delta t \left( -a_{\rm eff}\epsilon + \sqrt{\langle\phi(t)^2\rangle} z(t)\right) > 0.
\end{equation}
This implies that
\begin{equation}
    z(t) > \frac{(1 + a_{\rm eff}\Delta t)\epsilon}{\Delta t \sqrt{\langle\phi(t)^2\rangle}}
    \approx \frac{\epsilon}{\Delta t \sqrt{\langle\phi(t)^2\rangle}}.
\end{equation}
Averaging over the Gaussian distribution of $z(t)$, the average transition rate becomes
\begin{equation}
    R(\epsilon) = \frac{P(-\epsilon)}{\Delta t}
    \int^\infty_{\epsilon/\left(\Delta t \sqrt{\langle\phi(t)^2\rangle}\right)} Dz.
\end{equation}
Noting that $P(-\epsilon) \approx P(0)$ \mw{and integrating over non-negative values of $\epsilon$}, the zero-crossing rate (in positive or negative direction) becomes
\begin{equation}
    \mu(t) = \frac{P(0)}{\Delta t}
    \int_0^\infty Dz z \Delta t \sqrt{\langle\phi(t)^2\rangle}
    =\frac{\sqrt{\langle\phi(t)^2\rangle}}{\sqrt{2\pi}} P(0).
\end{equation}
\mw{The above equation is applicable to spin transitions from the negative to the positive side. The expression for transitions from positive to negative is identical.} 
Hence, we have shown that the zero-crossing rate is proportional to $P(0)$.
\zs{
\begin{equation}
    \langle\phi(t)^2\rangle
        \simeq
        \frac{2T}{\Delta t}
        +
        2J^2\xi^2Q(t,t) .
\end{equation}
The dominant term of the zero-crossing rate then becomes
 \begin{equation}
        \mu(t)
        \simeq
        \sqrt{\frac{T}
            {\pi\Delta t}}
        P(0).
\end{equation}
for small $\Delta t$ and non-zero $T$.}

\section{Numerical Experiments Setting}
\label{appendx: Experiments_Setting}

The annealing schedule for the temperature $T$ is defined as $T(\tau) = T_0 - (T_0 - T_e) \tau/S$, where $\tau = 1, 2, \dots$, and $S \mw{= 10^6}$ represents discrete time steps. The initial and final temperatures are set to $T_0 = 2$ and $T_e = 0.01$, respectively, while the drive parameter is fixed at $a = a_e$. For the linear annealing of $a$ (red curve), the parameter starts at $a_0 = -2$ and evolves to various final values, $a_e$, as shown in the different sub-figures. During this process, the temperature is held constant at $T = 0.01$. The annealing schedule for $a$ is given by $a(\tau) = a_0 - (a_0 - a_e)\tau/S$. For the constant parameter scenario, both $a$ and $T$ are fixed at $a = 2$ and $T = 0.01$, respectively, over the same time scale.

    {All experiments were performed independently 20 times using the Euler–Maruyama method with a time step of $\mw{\Delta t =} 0.05$, following the same annealing protocol for SK models with $N = 10{,}000$ spins.}

\begin{figure}[htb]
    \centering
    \includegraphics[width=0.45\textwidth]{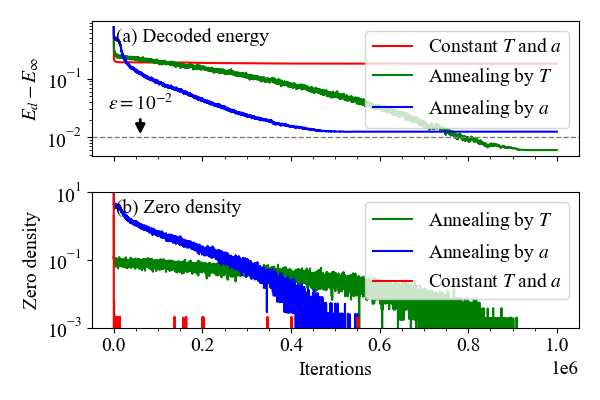}
    \caption{Trajectory of the decoded energy and $P(0)$ from a single CIM experiment with $a_e = 2$. Due to the results being numerically sampled from a single trial, the curve shown exhibits fluctuations.}
    \label{Fig.single_experiment}
\end{figure}

The evolution trajectory of the decoded energy and its numerical value of $P(0)$ for a single experiment is shown in Fig.~\ref{Fig.single_experiment}. As predicted, the zero-crossing rate for annealing by $a$ quickly drops to zero, causing the decoded energy to plateau. In contrast, annealing by $T$ delays the emergence of the effective gap, allowing it to gradually suppress other strategies and achieve the best performance.

\section{Numerical verification of effective gaps} \label{appendix:num}

\begin{figure}[htb]
    \centering
    \includegraphics[width=0.5\textwidth]{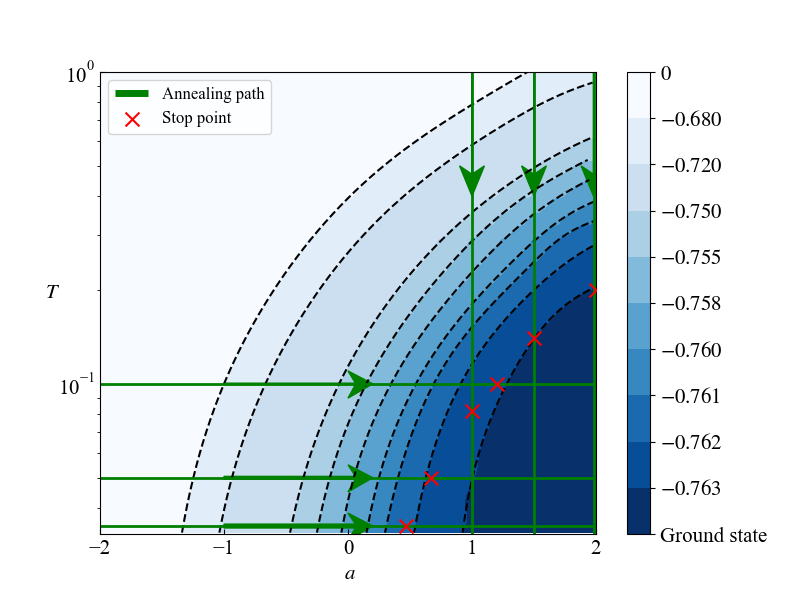}
    \caption{Numerical verification of the effective gap. The green lines show the annealing trajectories for both \(a\)-annealing and \(T\)-annealing, using the protocols described in Appendix~\ref{appendx: Experiments_Setting}. Red crosses mark representative points where the near-zero density \(P(0)\) vanishes numerically for systems of size \(N=10000\).  }
    \label{Fig:MC}
\end{figure}

\zs{ According to the schedule protocols discussed above, and as illustrated in Fig.~\ref{Fig.single_experiment}, one can estimate the parameter region in which the zero-crossing rate vanishes. We performed finite-size simulations with \(N=10000\) along several annealing trajectories and marked the points where the zero-crossing rate becomes statistically \mw{indistinguishable from} zero. These points define a numerical dynamical barrier, shown in Fig.~\ref{Fig:MC}. Specifically, we considered \(a\)-annealing from \(a=-2\) to \(a=2\) at fixed temperatures \(T=0.1\), \(0.05\), and \(0.035\), and \(T\)-annealing from \(T=1\) to \(T=0\) at fixed gains \(a=1\), \(1.5\), and \(2\). The numerically obtained dynamical barrier is consistent with the theoretical zero-density contours shown in Fig.~\ref{Fig.zero_density}.}

\zs{Using this dynamical barrier as a guide for trajectory design, we note that temperature annealing alone is not sufficient when the gain \mw{$a$} is too small, because the system then lacks a strong enough deterministic drive toward low-energy configurations. Conversely, gain annealing can benefit from enhanced exploration at sufficiently large \mw{$T$}. However, this strategy requires a delicate choice of temperature: if \mw{$T$} is too large, thermal fluctuations can wash out the energy landscape of the original Ising problem.}

\section{Gset Experiments}
\label{appendix:gset}
\mw{The concepts of energy barrier and related strategies, such as temperature annealing, are applicable beyond the SK model; we observe similar phenomena in general benchmarks and datasets. Here, we use the G1 and G6 instance from the popular Max-Cut benchmark dataset Gset~\cite{Ye_2003} as examples.}

\begin{figure}[htb]
    \centering
    \includegraphics[width=0.5\textwidth]{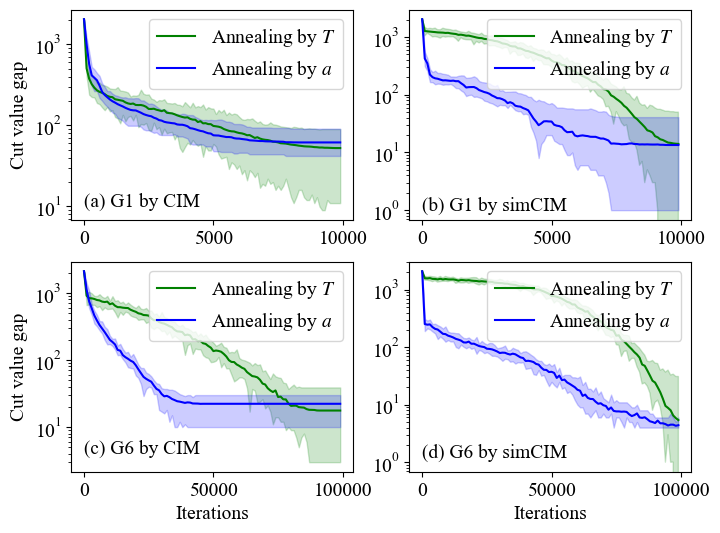}
    \caption{Comparison of cut values profiles obtained via different parameter-tuning strategies for CIM and simCIM on G1 and G6. The solid lines represent the average over 20 independent trials, while the shaded areas indicate the range between the maximum and minimum observed values. The $y$-axis shows the difference between the obtained energy and the known best solution. }
    \label{Fig:G1}
\end{figure}

\zs{As shown in Fig. \ref{Fig:G1}, similar to the SK model, temperature annealing in the CIM benefits from the delayed appearance of the dynamical barrier, resulting in better performance. In contrast, for simCIM, which benefits from the exposure of the optimal point, both temperature annealing and linear gain ($a$) annealing yield the same mean energy, outperforming the standard CIM. Notably, although the mean values for simCIM are identical, temperature annealing exhibits larger fluctuations. This leads to worse maximum values but better minimum values (\mw{increasing the chance of obtaining} the exact solution on G1 and G6). This potentially suggests that in practice, given computational resources for multiple trials and post-selection of the best result, temperature annealing remains the superior choice.}

\mw{The physical picture accounting for the above observations is the same as that in the SK model. For CIM, this can be seen from the dynamical equation of the individual nodes. While different network structures and optimization objectives provide different external driving terms, the combination of the linear gain and the cubic saturation in the internal driving term algebraically creates bifurcated spin values, resulting in the soft-hard dichotomy and the effective-gap picture. Indeed, simulation results for G1 in Supplemental Material of \cite{Zhou_2025} showed that the gap-opened phase exists in G1.}

\mw{For simCIM, due to the absence of the cubic saturation term, the linear gain term directly drives the spins towards the threshold values for positive gains, again resulting in the soft-hard dichotomy and the effective-gap picture. Furthermore, this direct drive leads to the simultaneous appearance of the gapless and binary phases, and hence the exposure of the optimal state as described in Section VI. These are also confirmed by simulation results for G1 in Supplemental Material of \cite{Zhou_2025}.}

\mw{We note, however, that this should be interpreted as an empirical observation rather than as a rigorous extension of the SK analysis to general sparse or structured graphs. In particular, G1 and G6 are not sparse in the asymptotic sense relevant here: the probability of a nonzero edge does not scale as $O(1/N)$, but remains of order $O(1)$. The case for sparse or structured Max-Cut instances remains to be explored.}

\zs{We also observe that the spectra of G1 and G6 approximately follow the semicircle law after normalization, making their large-scale spectral properties closer to those of the SK coupling matrix. This provides a possible universality-based explanation for the observed agreement, but we do not regard it as evidence for exact universality across general sparse or structured Max-Cut instances.}

\section{DigCIM Results}
\label{appendix:DigCIM Results}

\begin{figure}[tbh]
    \centering
    \includegraphics[width=0.5\textwidth]{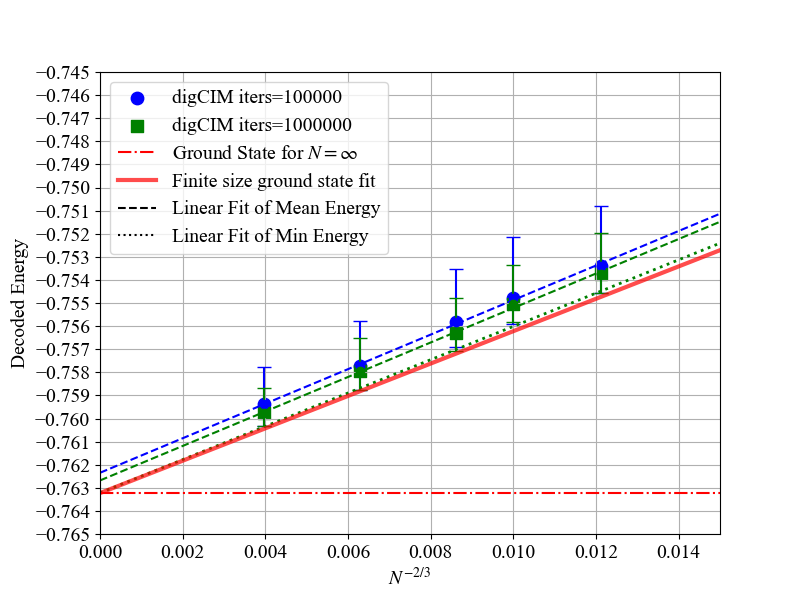}
    \caption{Energies obtained by digCIM using the Euler–Maruyama method (step size 0.1) with 100,000 and 1,000,000 iterations. It sets $a = -5$, with temperature annealing from 0.5 to 0. For each system size, N = 750, 1000, 1250, 2000, and 4000, we evaluate 200 independent SK instances and perform 30 runs per instance. Data points report the minimum decoded energy averaged over the 30 runs and then averaged across instances. Error bars show, for each instance, the mean of the minimum and maximum decoded energies observed over the 30 runs. The red line denotes the mean ground-state energy with finite-size corrections reported in Ref.~\cite{Boettcher2010}. The green dotted line indicates the fit for the expected minimum value after $10^6$ iterations, yielding an extrapolated infinite-size value of $-0.763203 \pm 0.00017$.}
    \label{Fig.dig_scale}
\end{figure}
DigCIM takes a form similar to simCIM:

\begin{equation}
    \frac{dx_i}{dt} =  a \varphi(x_{i})+ \xi \sum_{j=1, j \neq i}^{N} J_{ij } \sgn x_{j} + \zeta_i.
    \label{eq:digCIM}
\end{equation}
In our previous work~\cite{Zhou_2025}, we showed that this dynamics attains the SK ground state uniformly for any negative $a$. It also exhibits superior dynamical properties. As shown in Fig.~\ref{Fig.dig_scale}, it also obeys the finite-size scaling relation as in Eq.~(\ref{eq:finite_scaling}). Similarly, in Fig.~\ref{Fig.dig_sim_iter}, we plot the
average and minimum values of $E_{\rm dec}(t, \infty)$ versus $t$, confirming the scaling relations in Eq. (\ref{eq:power_law}), where $E_{\infty,\rm ave} = -0.76245 \pm 0.00018$, $B_{\rm ave} = 3.644$, $\xi_{\rm ave} =0.879$, $E_{\infty,\rm min} = -0.76317 \pm 0.00015$, $B_{\rm min} = 3.641$, and $\xi_{\rm min} =0.934$. Hence, applying the scaling relations
enables us to compute the ground state energy of the Ising model at the thermodynamic limit even more accurate than simCIM.

Comparing the solution quality of simCIM and digCIM under limited computational resources in Fig.~\ref{Fig.dig_sim_iter}, we see that, although they share the same asymptotic time complexity, digCIM benefits from a superior proportionality constant of the power law, resulting in a faster rate of energy decay.

\section{Soft-Spin Activity in Advanced Dynamical Solvers}
\label{appendix:advanced_solvers}

\zs{There are several popular dynamical solvers with more sophisticated mechanisms than the gradient-based systems discussed above, and many of them have demonstrated promising performance. Although a rigorous theoretical analysis of these solvers is beyond the scope of this work, our framework can still be used to numerically reveal common dynamical features and provide a brief qualitative discussion.}

\begin{figure}[htb]
    \centering
    \includegraphics[width=0.5\textwidth]{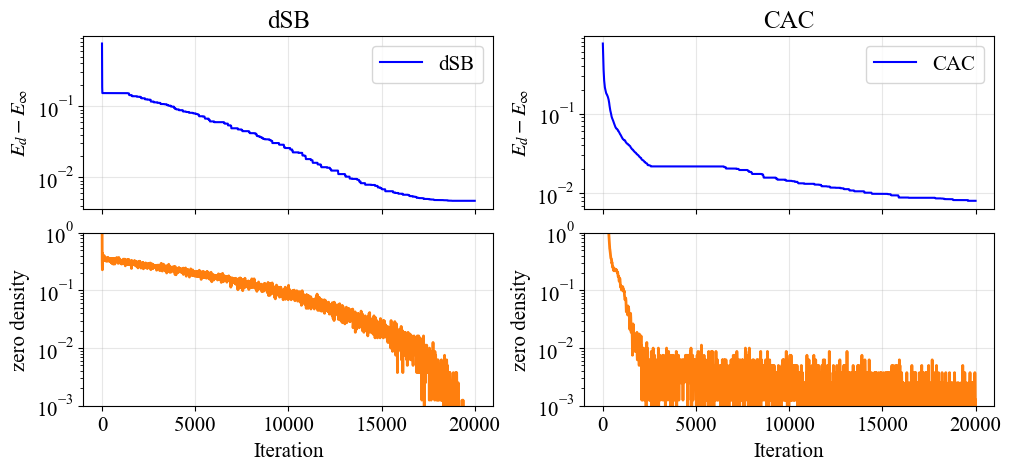}
    \caption{Trajectory of the decoded energy and \(P(0)\) from single-run experiments using dSB \mw{(discretized SBM)} and CAC. The parameter settings for dSB follow the recommendations of Ref.~\cite{Goto2021}, while those for CAC follow Ref.~\cite{Leleu2021-cz}.}\label{Fig.dsb_csc}
\end{figure}

\zs{One example is the simulated bifurcation machine (SBM) introduced in Ref.~\cite{Goto2021}. SBM evolves the dynamics by introducing momentum, which effectively plays a role analogous to temperature: it helps the system \mw{to} continue exploring \mw{the} configuration space by maintaining the effective gap and preserving a large population of soft spins until late in the evolution. Since SBM is based on a Hamiltonian structure similar to the systems discussed in the main text, its dynamics is expected to exhibit related phenomena. This expectation is supported by the similarity between the temperature-annealing trajectory in Fig.~\ref{Fig.single_experiment} and the SBM results shown in Fig.~\ref{Fig.dsb_csc}.}

\zs{Another example is the chaotic amplitude control (CAC) solver introduced in Ref.~\cite{Leleu_2019}. CAC dynamically suppresses amplitude heterogeneity and destabilizes shallow local minima by introducing an additional error dynamics that modulates the effective coupling strength $\xi$. From the perspective of the present framework, this destabilization mechanism can also maintain persistent spin flipping and preserve a finite soft-spin population as shown in Fig.~\ref{Fig.dsb_csc}, \mw{albeit} through a mechanism different from temperature annealing or momentum-induced exploration. Thus, while CAC is not explicitly described by the gradient-based DMFT equations analyzed in the main text, its empirical success is consistent with the broader picture that robust optimization requires avoiding premature hardening of the spin amplitudes and maintaining dynamical activity near the origin.}

\section{A Related Work}
\label{appendix:related_work}
\mw{During the finalization of this manuscript, we became aware of a parallel work analyzing the CIM's energy landscape for the SK model with ferromagnetic coupling via the laser gain parameter $a$ \cite{ghimenti2025geometry}. That study provided the geometry and dynamics of annealed CIM optimization in SK-related models with
hidden or planted solutions. It emphasized the geometric structure of the annealed landscape, particularly in
planted models.}

\mw{They showed that low-energy global minima develop soft modes that can guide optimization dynamics, and that these minima eventually become rigid, terminating further optimization gains from gain annealing. By contrast, the present work focuses on the unplanted SK optimization problem and develops a DMFT framework for the decoded Ising energy itself. In particular, our main practical conclusion is that temperature-only annealing can outperform conventional gain annealing in CIM/digCIM for a fixed deep target point in phase space.}

\mw{Nevertheless, the two analyses are consistent. Their observation that useful global minima possess soft modes, but eventually become rigid under further annealing, is closely related to our finding that optimization requires maintaining a finite density of near-zero, responsive soft spins. In our terminology, the loss of such soft degrees of freedom corresponds to entering the effective-gap region, where zero-crossing activity is suppressed and decoded-energy improvement stalls. Thus, their landscape-level picture of soft modes and rigidity provides a complementary geometric interpretation of the dynamical bottleneck identified here.}

\end{document}